\documentclass[noshowpacs,preprint,amsmath,superscriptaddress,pre]{revtex4}
\usepackage[dvips]{graphicx}
\usepackage{tabularx}
\usepackage{epsfig}
\usepackage{color}
\usepackage{subfigure}
\usepackage{xcolor}
\usepackage{soul}
\usepackage{amsmath,amssymb}
\usepackage{bm}

\begin{document}
\title{From random walks on networks to nonlinear diffusion}
\author{Carles Falc\'o}
\email{falcoigandia@maths.ox.ac.uk}
\affiliation{Mathematical Institute, University of Oxford, OX2 6GG Oxford, United Kingdom}

\begin{abstract}
Mathematical models of motility are often based on random-walk descriptions of discrete individuals that can move according to certain rules. It is usually the case that large masses concentrated in small regions of space have a great impact on the collective movement of the group. For this reason, many models in mathematical biology have incorporated crowding effects and managed to understand their implications. Here, we build on a previously developed framework for random walks on networks
to show that in the continuum limit, the underlying stochastic process can be identified with a diffusion partial differential equation. The diffusion coefficient of the emerging equation is in general density-dependent, and can be directly related to the transition probabilities of the random walk. Moreover, the relaxation time of the stochastic process is directly linked to the diffusion coefficient and also to the network structure, as it usually happens in the case of linear diffusion. As a specific example, we study the equivalent of a porous-medium type equation on networks, which shows similar properties to its continuum equivalent. For this equation, self-similar solutions on a lattice and on homogeneous trees can be found, showing finite speed of propagation in contrast to commonly used linear diffusion equations. These findings also provide  insights into reaction-diffusion systems with general diffusion operators, which have appeared recently in some applications.

\end{abstract}

\date{\today}

\maketitle

\section{Introduction}

Interactions between individuals are fundamental in order to decipher collective motility rules.
This process is relevant in many fields, from the movement of cells during development and tumour formation \cite{ArmstrongPainterSherratt}, to the spreading of diseases or opinions \cite{ishii2019opinion,AsllaniNonlinearRWEpidemics}. One of the simplest forms of interaction takes place when individuals compete for available space, which is usually referred to as \emph{volume exclusion}. Such a basic mechanism can have drastic changes in model dynamics and usually comes in the form of localized repulsive forces between individuals that are close to each other \cite{AlertTrepat,carrilloSciannaColombi,CarrilloMurakawaCellAdhesion}. Crowding effects play an important role in biology across different scales \cite{rivera2011VolumeExclusion,minton2000implications,Trush5861,murakawa2015continuous} but also make sense in other contexts, where the behaviour of single individuals is highly influenced by large masses. Recent models on networks have focused on incorporating this \emph{volume-filling} mechanism \cite{WilsonWoodhouseBaker,AsllaniCrowding,CarlettiNonlinear}, but there is also a vast literature coming from the field of mathematical biology that at its core, is motivated by the same basic ideas -- to name a few \cite{DysonMacroscopicCrowding,PillayVolumeExclusion,DysonVolumeExclusion,WilsonBruna,BrunaChapman,KhainAdhesive,SimpsonAdhesive}.

Many of these models, however, focus on continuum descriptions of the system of interest, which are based on some kind of diffusion partial differential equation (PDE). While the case of linear diffusion has probably been the most popular choice, this is not suitable when there is volume exclusion because it represents purely random motion. Probably, one of the first alternatives used to model biological populations \cite{skellam} was the so-called porous-medium equation (PME), introduced in this context in \cite{Gurtin1977OnDiffusion}.
The PME is a nonlinear diffusion equation (see \cite{PME} for an exhaustive and detailed analysis), where the diffusion coefficient increases as a power-law function of the density so that
\begin{equation}
    \frac{\partial \rho}{\partial t} = \Delta \rho^m = m\nabla\cdot\left(\rho^{m-1}\nabla\rho\right).
    \label{PME}
\end{equation}
Note that $m = 1$ corresponds to the usual linear diffusion. The cases $m>1,\,m<1$ are usually called \emph{slow diffusion} and \emph{fast diffusion} because of how quickly their solutions decay in time compared to the reference case given by linear diffusion.

Different arguments can be found in the literature to derive the PME. The case of exponent two is particularly interesting as it can be identified with particle-based models \cite{gurney2, DysonMacroscopicCrowding}, and also with on-lattice models where agents have different aspect ratios \cite{bakerAspectRation}. In fact the exponent of the PME has been related to the size of the agents \cite{bakerAspectRatio2}. Moreover, the $m = 2$ case is also closely related to statistical mechanics, as it emerges as the thermodynamic limit of a system of interacting agents \cite{Oelschlger1990LargeSO,carrillo2019aggregation,CarrilloMurakawaCellAdhesion}. The PME with $m = 3$ has also appeared in applications, and has been suggested as \emph{the simplest model} to relate the dispersal velocity to both the density and its gradient \cite{TopazBertozziLewis}.

Understanding the interplay between diffusion and crowding effects on networks is also relevant, as many processes are more suitably described  by spatially discrete models. This raises the question of whether one can translate some of the mentioned PDEs to the graphs setting \cite{MaasPME,EspositoNonlocal,MarkusArxiv}. Moreover, continuum models are usually not only limited to diffusion but include other type of mechanisms or \emph{reactions}, and these have also appeared in the discrete case \cite{AsllaniTuringPatterns,FisherKPPGoriely}.

Here, we build on a graph random-walk framework to show that on regular lattices, one can identify a continuous limit for the underlying stochastic process. The emerging macroscopic PDE has the form of a general diffusion equation, with a density-dependent diffusion coefficient which is directly related to the transition probabilities of the random walk. This \emph{discrete-to-continuum} approach provides a direct link between models at different scales and gives new insights on the process at the microscopic level. In particular, we show that for porous-medium type equations one also observes mass propagation with finite speed on networks, in contrast with the commonly used linear diffusion equation. The propagation speed can be  found analytically for simple graphs. On finite networks, we also fully characterize the stationary distribution of the stochastic process and the relaxation time needed to reach this state, which depends on the network structure and on the diffusion coefficient. These analytical results can also be applied to reaction-diffusion systems, which we briefly discuss and compare to the recent findings of \cite{FisherKPPGoriely}.

\section{Preliminaries}

We build on \cite{AsllaniCrowding,CarlettiNonlinear} and consider $N$ agents moving on a network with $M$ nodes -- for simplicity we consider unweighted undirected graphs. A novel feature of these works is that each node has a finite carrying capacity $K$ which sets an upper limit for the number of agents that can be at the same node. Here we assume that all nodes have the same carrying capacity, but in principle it could also be node-dependent -- see \cite{AsllaniCrowding}. We denote by $A_{ij}$ and $k_i$ the elements of the adjacency matrix and the degree of each node, respectively. We also represent the state of the process via a vector $\mathbf{n}=\left(n_1,\ldots,n_M\right)$, where $0\leq n_i\leq K$ is the number of agents at node $i$. \textcolor{black}{In \cite{CarlettiNonlinear,AsllaniCrowding} the agents perform a node-centric random walk only allowing for transitions to neighbouring nodes, for which $A_{ij} = 1$. Considering such a process, where the hopping rate is independent of the node degree, usually gives place to a diffusion operator involving the random-walk normalized Laplacian matrix, $L'$, and to stationary states which depend on the degree of each node, $k_i$ \cite{LambiottePorterMasuda}. However, more recently it has been argued that the edge-centric choice, where the hopping rate is also proportional to the node degree, is more suitable for applications where one would expect uniform distributions in the infinite-time limit \cite{FisherKPPGoriely}. In this work, and following this discussion, we focus on the edge-centric case, although the node-centric case is studied for a specific model in Appendix \ref{appendix:edge_centric}.}

We further assume that the transition probabilities of the random walk are modulated according to the local density of agents. More specifically, if there is an agent in node $i$, we assume that the \emph{willingness} to leave such node is given by a function $f$, which depends on the local density, $n_i/K$; and the \emph{attractiveness} of a neighbouring node $j$ is given by another function $g$ which depends on the density at this node, $n_j/K$. These functions then measure the influence of crowding on motility \cite{li2022interpreting}. Note that this framework is different from other biased random-walks on networks where the bias depends on the network topology rather than on how crowded a particular node is \cite{LatoraEntropyRate,FronczakBiasedRW,SinatraMaximalEntropy,BonaventuraBiasedRW}.

In particular, the stochastic process is given by the transition probabilites $T(\mathbf{n}'|\mathbf{n})$ from state $\mathbf{n}$ to state $\mathbf{n}'$, and the master equation for the evolution of the probability of state $\mathbf{n}$ at time $t$
\textcolor{black}{
\begin{equation}
    \frac{\mathrm{d}}{\mathrm{d}t}P(\mathbf{n},t) = \sum_{\mathbf{n}'}\left[T(\mathbf{n}|\mathbf{n}')P(\mathbf{n}',t)-T(\mathbf{n}'|\mathbf{n})P(\mathbf{n},t)\right].
    \label{eq:master_equation}
\end{equation}}
Since exchanges of particles only happen between pairs of nodes $(i,j)$ for which $A_{ij} = 1$, and only increments and decrements of one unity per unit time in the number of particles are allowed, the transition probabilities $T(n_i-1,n_j +1|n_i,n_j)$ are sufficient to characterize the process. According to our previous considerations these read\textcolor{black}{
\begin{equation}
    T(n_i-1,n_j +1|n_i,n_j) = \lambda A_{ij}f\left(\frac{n_i}{K}\right)g\left(\frac{n_j}{K}\right).
    \label{eq:Transition_edge_centric}
\end{equation}
Here $\lambda$ is just a hopping rate giving the timescale of the process.
Usually, one thinks of $f$ and $g$ as non-decreasing and non-increasing functions, respectively, as hops from low density regions to crowded nodes are less likely to occur.}

\textcolor{black}{In the limit of large $K$ and for small densities, we can expect the transition probabilities \eqref{eq:Transition_edge_centric} to be small and consequently, $P(\textbf{n},t)$ will be peaked around the macroscopic mean. In this situation, correlations can be neglected -- see \cite[Chapter X]{VanKampen} -- and it makes sense to consider the mean-field approximation $\langle f(\cdot)\rangle =f\left( \langle \cdot\rangle\right)$ and similarly for $g$. Under this approximation, $1/K$ corrections vanish,
and one can explicitly write an evolution equation for the mean-field node density $\rho_i(t) = \lim_{K\rightarrow\infty}\langle n_i\rangle/K$ 
-- see \cite{AsllaniCrowding,CarlettiNonlinear} for more details --
\begin{equation}
        \frac{\mathrm{d}\rho_i}{\mathrm{d}t} = -\lambda \sum_jL_{ij}\left(f(\rho_j)g(\rho_i)-f(\rho_i)g(\rho_j)\right),\label{General_Model}
\end{equation}
where here $L_{ij} = \delta_{ij}k_i-A_{ij}$ is the usual graph Laplacian \cite{LambiottePorterMasuda}. The approximation is exact when $K\rightarrow\infty$ \cite{McKaneNewman1,McKaneNewman2}.}

\textcolor{black}{Previous studies also support the validity of the mean-field approximation for finite $K$ \cite{SimpsonBakerMeanField2,RuthMeanFieldCorrections,MurrellUlf}. In general, one can expect the approximation to work in the absence of proliferation, and for small movement bias. Note here that jumps from a given node are biased if the densities of the neighbouring nodes are different and $g$ is not constant. Whenever the transition probabilities are dominated by $f$ instead of $g$, movement bias should be small.}

 One important property that follows from \eqref{General_Model} is the conservation of mass which will be used later to characterize the stationary distribution. We can then define the average density as $\Bar{\rho} = M^{-1}\sum_{i = 1}^M\rho_i$, which we can calculate from the initial condition and does not depend on time.

Observe that in the case $f(x) = x$ and $g(x) = 1$, the linear diffusion equation
is recovered. In this work we consider general forms of $f$ and $g$, but for some applications we look at the case given by a specific choice of these functions representing only local repulsive interactions. In particular we will study the case $f(x) = x^m$ and $g(x) = 1$, meaning that the \emph{attractiveness} of any node is independent of its density, and that the \emph{willingness} to leave a node is modulated by the exponent $m\geq 0$. As we will discuss in the next sections this choice of transition probabilities is closely related to the macroscopic PME.

Note too that a simplified version of the Kuramoto model on networks \cite{arenas2008synchronization} is recovered for $f(x) = \sin x$ and $g(x) = \cos x$. In this case, Eq. \eqref{eq:Transition_edge_centric} should be reinterpreted as $\rho_i$ is no longer a density but a phase which could be negative.

Several other choices of $f$ and $g$ have been studied before. This framework was initially derived in \cite{AsllaniCrowding}, in the node-centric setting, with $f(x) = x,\,g(x) = 1-x$ to solve the inverse problem of finding the connectivity distribution of a network. Note however that this choice in the edge-centric setting case \eqref{General_Model}, results in linear diffusion. Later, it was extended to allow for nonlinearities of the type $f(x) = x, \,g(x) = (1-x)^\sigma$ in \cite{CarlettiNonlinear}. Thus, we remark that the choice $f(x) = x^m$ is novel and has not been studied before -- except for its stationary distribution in the node-centric setting, in \cite{CarlettiNonlinear}.

\section{Macroscopic limits}
\label{macro_limit}

Before moving to study the properties of Eq. \eqref{General_Model} we analyze it in the case of \textcolor{black}{regular lattices}, where the distance between nodes is given by $a$. In fact, the approach taken here is similar to the coarse-graining method used to link continuous-time discrete-space equations with macroscopic models \cite{painter2002volume}. We imagine that the process takes place on a square grid graph, where the degree $k$ only depends on the dimension of the grid. We assume for simplicity that the grid is one-dimensional and thus $k =2$ and $A_{ij} = \delta_{i,i+1} + \delta_{i,i-1}$. Note however, that the following results generalize easily for any higher-dimensional square grid. Then Eq. \eqref{General_Model} simplifies to
\textcolor{black}{
\begin{eqnarray}
      \frac{\mathrm{d}\rho_i}{\mathrm{d}t} =\lambda\big[g(\rho_i)\left(f(\rho_{i+1})+f(\rho_{i-1})\right)
      -f(\rho_i)\left(g(\rho_{i+1})+g(\rho_{i-1})\right)\big].\label{1d_lattice_general}
\end{eqnarray}}
We take a continuous approximation of the one-dimensional lattice in the limit where the number of nodes $M$ tends to infinity, and the distance between two consecutive nodes, $a$, tends to zero. We identify $\rho(x,t) = \rho_i(t)$ (also $\mathrm{d}\rho_i/\mathrm{d}t = \partial_t \rho$) and $\rho(x\pm a,t) = \rho_{i\pm 1}(t)$. By expanding $\rho_{i\pm 1}$ as a Taylor series for small values of the distance between nodes $a$, we obtain
\begin{equation}
    \rho_{i\pm 1} =  \rho \pm a\partial_x\rho+\frac{a^2}{2}\partial_{xx}\rho+ o(a^2),
\end{equation}
and again for $f(\rho_{i\pm 1})$ (and similarly for $g(\rho_{i\pm 1})$),
\begin{eqnarray}
      f(\rho_{i\pm 1}) = f(\rho)\pm af'(\rho)\partial_x\rho + \frac{a^2}{2}f'(\rho)\partial_{xx}\rho
      +
      \frac{a^2}{2}(\partial_x \rho)^2 f''(\rho) + o(a^2).
\end{eqnarray}
Note that this last expansion is in fact exact whenever $f$ and $g$ are polynomials in $\rho$. By using this, and writing \textcolor{black}{\begin{equation}h = f'g-fg',\end{equation}}Eq. \eqref{1d_lattice_general} reduces to
\textcolor{black}{
\begin{equation}
    \partial_t\rho = \lambda a^2\left[h(\rho)\partial_{xx}\rho + (\partial_x\rho)^2h'(\rho)\right]+o(a^2).
\end{equation}}
Taking now the macroscopic limit $a\rightarrow 0$ alongside the scaling $\lambda = O(a^{-2})$ we obtain
\begin{equation}
     \partial_t\rho =h(\rho)\partial_{xx}\rho + (\partial_x\rho)^2h'(\rho) =\partial_x\left(h(\rho)\partial_x \rho\right) = \partial_{xx} H(\rho).\end{equation}
where $H' = h$.
Note that without loss of generality we assumed here that $\textcolor{black}{\lambda a^2\rightarrow 1}$ as $a\rightarrow 0$. As mentioned before, this same argument holds in square grid graphs of higher dimensions and gives rise to the nonlinear diffusion equation
\begin{equation}
    \partial_t \rho = \Delta H(\rho) = \nabla\cdot\left(h(\rho)\nabla\rho\right),
    \label{eq:diffusion_limit}
\end{equation}
with a density-dependent diffusion coefficient $h(\rho)$. Note that for well-posedeness of the equation one needs $h\geq 0$, which is always the case given that $f$ is non-decreasing and $g$ is non-increasing. In the case where these assumptions fail -- e.g. $g$ is non-monotonic due to the possible presence of an aggreggation mechanism \cite{nonlinearRWEcology} -- the macroscopic PDE could be ill-posed. However, in such case, it still makes sense to analyze \eqref{General_Model}, which is a discrete model -- a similar phenomenon is studied in \cite{negativeDiffusionAnguige}.

In the particular case where $H(\rho) = \rho^m$ we obtain the PME \eqref{PME}. This can be achieved for example when the transition probabilities only depend on local densities $f(x) = x^m$ and $g(x) =1$, which represents a process with very localized repulsive interactions (but there are other possibilities). In this case the diffusion coefficient is
\begin{equation}
    h(\rho) = m\rho^{m-1}.
    \label{Diffusion_PME}
\end{equation}
Moreover, Eq. \eqref{General_Model} can be written in such a way that reminds one of the macroscopic PME
\begin{equation}
    \frac{\mathrm{d}\rho}{\mathrm{d}t} = -\lambda  L \rho^m.
    \label{NPME}
\end{equation}

We note here that different choices of $f$ and $g$ can yield macroscopic equations which are well-known PDEs. For example, it is instructive to analyze the case studied in \cite{CarlettiNonlinear} $f(x) = x$ and $g(x) = (1-x)^{\sigma}$, with $\sigma \geq 0$. This yields the degenerate diffusion coefficient
\begin{equation}
    h(\rho) = (1-\rho)^{\sigma-1} \left[1+(\sigma-1)\rho\right].
    \label{eq:diffusion_mobility}
\end{equation}
Similar mobility coefficients in PDEs have appeared in mathematical biology to prevent overcrowding \cite{BurgerKellerSegel,painter2002volume,calvezCarrillo} but also in mathematical physics to describe fermionic systems where particles obey the exclusion principle \cite{KaniadakisFermions,CarrilloRosado} -- these usually correspond to the case $\sigma = 2$. Under the substitution $(1-\rho)^{\sigma-1}\mapsto(1+\rho)^{\sigma-1}$ we obtain instead an equation describing a system of bosons under relaxation \cite{CarrilloRodrigo}.  Moreover, such equations can also be derived from microscopic rules following a probabilistic argument -- see \cite{burgerSchlakeWolfram} and also \cite{BurgerDiFrancescoMobilityLattice} for a case with two species.
Here, $h(\rho)$ has  completely different behaviours in the cases $\sigma =1$ , when this reduces to linear diffusion; $\sigma > 1$, when $h(\rho) < 1$ and is a decreasing function; and $\sigma < 1$, which makes $h(\rho)>1$ increasing and unbounded. This could be directly related to previous works studying the exploration efficiency of different random-walk models on networks \cite{CarlettiNonlinear,LatoraEntropyRate}. In fact, and as we will see, the diffusion coefficient is not only related to the macroscopic PDE but also determines the timescale of the process on finite graphs.

\section{Self-similar solutions on infinite graphs}

Many PDEs of the form of \eqref{eq:diffusion_limit} present well-known solutions in the case of infinite domains. Of particular interest is the case of the PME, with $h(\rho) = m\rho^{m-1}$, which solutions in the real line tend to the so-called Barenblatt profiles
\cite{carrillo2000asymptotic}. In one spatial dimension, these can be found by looking for self-similar solutions of the form $\rho(x,t) = t^{-\beta}F(xt^{-\beta})$ and give place to
\begin{equation}
    \beta = \frac{1}{m+1},\quad F(y) = (C-\kappa y^2)_+^{1/(m-1)},
    \label{eq:barenblatt}
\end{equation}
where $(y)_+ = \max(y,0)$, and $C$ and $\kappa$ are constants related to the exponent $m$ and the total mass \cite{PME}. Note that according to this relation, mass is propagated following the relation $x \sim t^\beta$. In the limit $m\rightarrow 1$ we obtain the heat kernel solution for the linear diffusion PDE. Here we analyze, following a self-similarity argument, a porous-medium type equation  \eqref{NPME} on two infinite graphs, namely a regular lattice and a $q$-homogeneous tree. We find that in contrast with the case of linear diffusion, mass propagation occurs at a finite speed. These calculations give also insights into nonlinear diffusion on more general graphs, where the observed solutions resemble those of the macroscopic PME.

\subsection{Regular lattice}

In the case of an infinite regular lattice we show that solutions of Eq. \eqref{NPME} agree with those of the macroscopic PME and in the long-time limit tend to the Barenblatt profile \eqref{eq:barenblatt}. On this domain, Eq. \eqref{NPME} simply reads
\begin{equation}
    \frac{\mathrm{d}\rho_i}{\mathrm{d}t} = \rho_{i-1}^m + \rho_{i+1}^m - 2\rho_i^m.
\end{equation}
We assume that initially, all the mass is placed at $i=0$ and we allow for integer values of $i$.
Now assume the self-similar ansatz $\rho_i = t^{-\beta} F(y)$ with $y =it^{-\beta}$, and substitute into the above expression to find
\begin{equation}
    -\beta t^{-\beta-1}\left[F(y)+yF'(y)\right] = t^{-m\beta}\left[F(y - t^{-\beta})^m+F(y + t^{-\beta})^m+F(y)^m\right],
\end{equation}
where the prime denotes differentiation with respect to $y$.
By expanding the right-hand side for large $t$, and eliminating the time dependence, we find $\beta = 1/(m+1)$, and also the differential equation for $F$
\begin{equation}
    (m+1)\left(F(y)^m\right)^{''} + yF'(y)+F(y) = 0.
\end{equation}
This equation is usually called the \emph{profile equation} \cite{PME} and can be integrated to obtain the Barenblatt profile \eqref{eq:barenblatt}.

Thus we have proved that on a lattice, \eqref{PME} behaves as its continuum limit, the PME.
\subsection{Homogeneous trees}

We now move to the case of infinite $q$-homogeneous trees, where each node has degree $q+1$. We assume $q>1$ since $q =1$ corresponds to the case of the lattice studied in the previous section. Now, the long-time behaviour is more complex, but it can be understood by looking at the continuum limit again. Eq. \eqref{NPME} reads now
\begin{equation}
     \frac{\mathrm{d}\rho_i}{\mathrm{d}t} = q\rho_{i-1}^m + \rho_{i+1}^m - (q+1)\rho_i^m,\label{eq:tree}
\end{equation}
where $\rho_i$ is representative of the set of nodes in the $i$th generation of the tree. Here we assume that initially, all the mass is located at nodes with $i = 0$, and that the tree extends for negative and positive values of $i$.

Following the approach in the previous section, one can now take the continuum limit of Eq. \eqref{eq:tree}. By doing so, we obtain the nonlinear diffusion PDE with convection
\begin{equation}
    \partial_t\rho = a^2(q+1)\partial_{xx}\rho^m - a(q-1)\partial_x\rho^m,
\end{equation}
where $a$ stands for the distance between nodes. Note that the equation can be explicitly solved for the $m =1$ case to obtain a travelling gaussian with speed $\sim (q-1)$. In fact, it is easy to see via a self-similar argument as in the previous section that this is also true for the discrete case \eqref{eq:tree}. We focus on the case $m>1$.

These type of equations with nonlinear diffusion as well as convection, have been widely studied and depending on whether diffusion or convection dominates, we may find different long-time behaviours. When both diffusion and convection have the same exponent $m >1$, the dominant term is the convective one, and in the long-time limit, the diffusive part can be neglected \cite{laurencot_simondon_1998}. Asymptotic solutions then can be obtained from the equation $\partial_t \rho + \partial_x\rho^m = 0$ and follow  $\rho(x,t) \sim t^{-\gamma}G(z)$ with $z = xt^{-\gamma}$ and
\begin{equation}
    \gamma = \frac{1}{m},\quad G(z) =\begin{cases}z^{1/(m-1)},&\mbox{for }0<z<C_m,\\0&\mbox{otherwise,}\end{cases} \label{eq:tree_continuum}
\end{equation}
where the constant $C_m$ is determined by normalization.

Following these ideas we may now assume the self-similar ansatz $\rho_i = t^{-\gamma}G(z)$ with $z = it^{-\gamma}$ for Eq. \eqref{eq:tree}. By substituting into \eqref{eq:tree}, expanding again for large $t$, and eliminating time-dependence, we find $\gamma = 1/m$ and the differential equation for $G$
\begin{equation}
    m(q-1)\left(G(z)^m\right)' + zG(z) + G(z) = 0.
\end{equation}
By looking for solutions $G(z)\sim z^\delta$ we obtain $\delta = 1/(m-1)$ as expected from the continuum limit solution \eqref{eq:tree_continuum}.

Observe that in the discrete case we have obtained solutions that agree with the corresponding continuum limits, showing propagation speeds which heavily depend on the exponent $m$. It would also be interesting to study these equations with an added reaction term, as in the linear diffusion case these can show different behaviour with respect to the continuum PDE  \cite{HoffmanFisherKPP}.
We have also shown that the propagation speed of solutions  depends on the network structure, and while it may be possible to extend these results to similar graphs, the case of general complex networks seems challenging. However, for finite and connected graphs the situation is simpler as after some time all nodes present non-zero densities. We will see in the next sections that this allows us to characterize the timescale of the process.


\section{Stationary distribution and relaxation time}

From now on and unless stated otherwise, we assume that we work with finite and connected networks. As expected for an edge-centric random walk, the stationary distribution of Eq. \eqref{General_Model} for any choice of $f$ and $g$ is a uniform distribution, as long as $f$ is non-decreasing and $g$ is non-increasing. To see this note that a stationary state $\rho^*$ of \eqref{General_Model} satisfies
\begin{equation}
    \sum_j L_{ij}\left(f(\rho_j^*)g(\rho_i^*)-f(\rho_i^*)g(\rho_j^*)\right) = 0,
\end{equation}
for all $i = 1,\ldots,M$. In particular this means that $v(i)$ defined by $v_j(i) = f(\rho_j^*)g(\rho_i^*)-f(\rho_i^*)g(\rho_j^*)$ is an eigenvector of $L$ with zero eigenvalue, for $i = 1,\ldots,M$. On a connected graph, $v_j(i)$ is a constant independent of $j$ \cite{chung1997spectral}, and setting $i = j$ we see that $v_j(i) = 0$.

    Note that here, we assumed that $f(\rho_i^*),\, g(\rho_i^*)\neq 0$, which is always true given that $f$ and $g$ are nonzero for positive density values. In cases where this assumption fails, steady states in general could depend on initial conditions and on the precise form of these functions. This is also the case in general when $f$ and $g$ fail to be monotonic, a situation that has appeared before in different settings \cite{nonlinearRWEcology,VanDerMeer,chaos_nonmonotonic}.

\textcolor{black}{
Reordering the terms gives $f(\rho^*_i)/g(\rho_i^*)=f(\rho^*_j)/g(\rho_j^*)$, which is met for any $i$ and $j$ only if both sides are equal to a constant, $c$. Hence the stationary distribution must satisfy the relation \begin{equation}
    \frac{f(\rho^*_i)}{g(\rho_i^*)} = \frac{f(\rho^*_j)}{g(\rho_j^*)} =c.
\end{equation}
This equation always has a unique solution given that $f$ is non-decreasing and $g$ is non-increasing, and from it, we deduce that $\rho^*$ is uniform as it cannot depend on $i$. Using the mass conservation property we finally obtain $\rho_i^*=\Bar{\rho}$ for every node, with $\Bar{\rho} = M^{-1}\sum_{i = 1}^M \rho_i$ being the average density that we can calculate from the initial condition and does not depend on time.}

\textcolor{black}{The main challenge when we look for the relaxation time of Eq. \eqref{General_Model} lies in the nonlinearities $f$ and $g$. However, in the edge-centric case, this can be done for general forms of these two functions. We seek for solutions of the form $\rho_i = \Bar{\rho} + \epsilon  \eta_i$ for $\epsilon \ll 1$. Since $\Bar{\rho} M=\sum_i\rho_i = \sum_i\rho_i^*$, we require the perturbation $\eta$ to have zero total mass. Bringing together the $O(\epsilon)$ terms we obtain
\begin{equation}
    \frac{\mathrm{d}{\eta_i}}{\mathrm{d}t}=-\lambda h(\Bar{\rho})\sum_j L_{ij}\left({\eta_j}-{\eta_i}\right), \label{eq:linearised}
\end{equation}
or more compactly ${\mathrm{d}\eta}/{\mathrm{d}t}=-\lambda h(\Bar{\rho}) L\eta.$
Note that this is the usual linear diffusion equation with a rescaled time. We know then from \cite{LambiottePorterMasuda} that the relaxation time to the stationary state is simply
\begin{equation}
    \tau^{-1}_1 = h(\Bar{\rho})\lambda \mu_1,\label{eq:relaxation_time}
\end{equation} with $\mu_1$ being the smallest nonzero eigenvalue of $L$. This expression relates the relaxation time of the random-walk with the diffusion coefficient appearing in the continuum limit, and thus provides a direct link between the microscopic nature of the process and the macroscopic dynamics. It also shows the same scaling as in the PDE \eqref{eq:diffusion_limit} up to a constant, which is given by the network structure.}

This derivation could lead us to think that the nonlinear process is somehow equivalent to its linearised version \eqref{eq:linearised} up to a constant given by the diffusion coefficient. However, and as we will see in the next section, this is not the case, and the timescale of the process does not fully determine its dynamics, which in general depend on the nonlinearities $f$ and $g$. The linear approximation is only valid once the mass has propagated to all nodes, similarly to what happens in the continuum case on finite domains and under no-flux boundary conditions. This also becomes evident when one thinks of the simplified Kuramoto model, which is recovered for $f(x) = \sin x,\, g(x) = \cos x$. In this case, we obtain $h = 1$, although its dynamics are different from the linear diffusion case, which also has $h = 1$.

The derived relaxation time, should be understood then in the sense of the asymptotic decay of the solutions in finite graphs. In other words, it provides an estimate of how $\rho$ tends to $\rho^*$. By projecting $\eta$ into a basis of eigenvectors of $L$, it is easy to see that in the long-time limit, $\eta\sim e^{-t/\tau}u_1$, where $u_1$ is the eigenvector of $L$ associated with $\mu_1$ \cite{LambiottePorterMasuda}. A direct measure of the convergence to the stationary solution is then given by the 2-norm of $\rho-\rho^*$, which in the long-time limit follows $\Vert\rho-\rho^*\Vert_2\sim e^{-t/\tau_1}$. Note however that we do not have this estimate on infinite graphs. In that case, finding the timescales of the process as in \cite{diaz2022time} seems challenging due to the nonlinearities in our equation.

\section{Slow and fast diffusion}

We continue the discussion on the previous section with special emphasis on the case given by Eq. \eqref{NPME}, which corresponds to the choice $h(\rho)=m{\rho}^{m-1}$.
As mentioned earlier, the regimes $m<1$, $m=1$, $m>1$ are usually referred to as \emph{fast diffusion}, \emph{linear diffusion}, and  \emph{slow diffusion} in the context of the PME \eqref{PME}. Given these known differences between the linear diffusion PDE and the PME with exponent $m\neq 1$, one could then wonder how solutions of Eq. \eqref{NPME} depend on $m$ for general graphs.

\begin{figure*}
    \centering
    \includegraphics[trim=100 0 100 0,width = 1\textwidth]{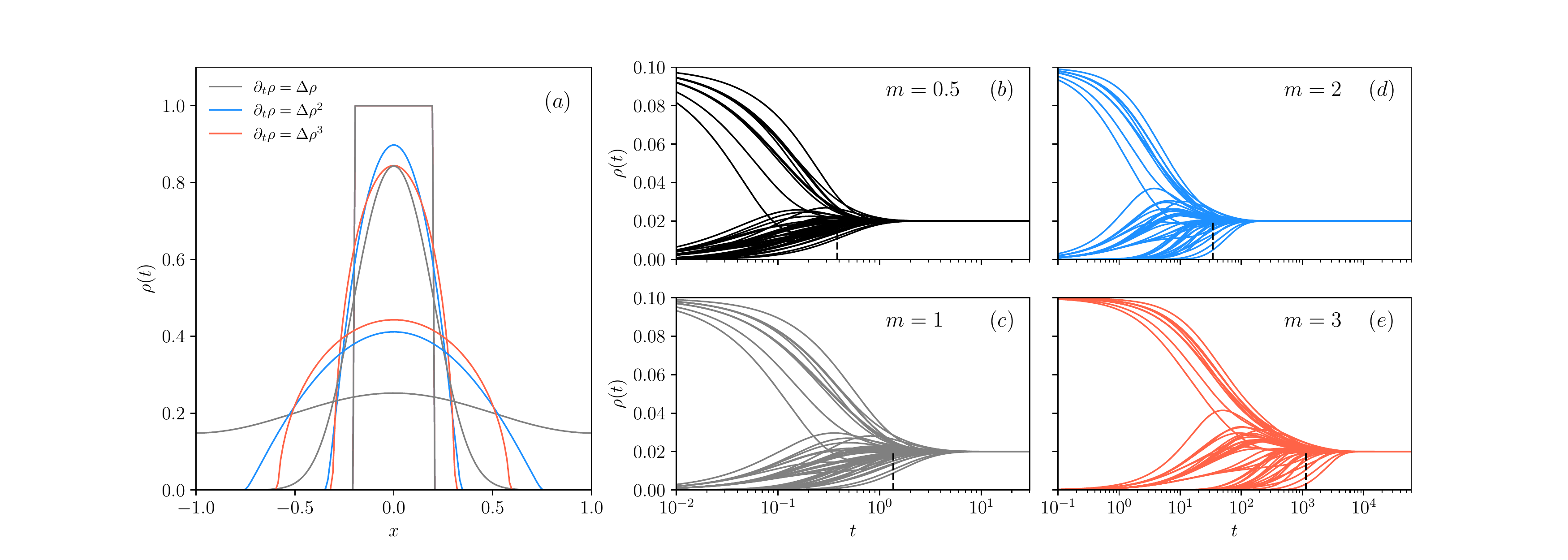}
    \caption{Nonlinear diffusion yields mass propagation at a finite speed, in contrast with linear diffusion. (a) PME in a box of length two with no-flux boundary conditions and for exponents $m=1,2,3$ -- $m=1$ corresponds to the linear diffusion equation. Solutions at three different time points starting from a compactly supported initial condition. Note the slower decay of the PME for exponents $m=2,3$. (b)-(e) Numerical solutions of Eq. \eqref{NPME} with $\lambda =1$, for an Erdos-Renyi network with $M = 50$ and a probability of drawing any possible edge $p = 0.1$. The four numerical simulations start from the same initial conditions and take place on the same network. At $t = 0$, 10 randomly selected nodes start with $\rho_i = 0.1$, so $\bar{\rho} = 1/M$. Dashed lines correspond to the relaxation time given by Eq. \eqref{eq:relaxation_time}.}
    \label{fig:1}
\end{figure*}

In fact, the situation on networks is similar to the case of the PDE \eqref{PME} under no-flux boundary conditions -- as shown in Fig. \ref{fig:1}a. In the PDE setting, for early times, one would expect the density to approach the solution in the whole space, but as soon as the density reaches the boundary, nonlinear and linear diffusion tend to the uniform steady state in a similar way.

As mentioned in previous sections, in the whole space -- where  there are no boundary effects -- and in the $m>1$ case, solutions to the PME \eqref{PME} with an initial point source are of the form of the so-called Barenblatt profiles \cite{PME}. These differ from the Gaussian solution to the linear diffusion equation as they are compactly supported, and have a power-law decay in time $t^{-\beta}$ with an exponent decreasing with $m$. This means that the speed of propagation is finite in contrast with the case of linear diffusion. In fact, the larger the exponent $m$, the slower the solution decays. This is depicted in Fig. \ref{fig:1}a, where we compare the PME solutions with different exponents to the Gaussian solutions of the linear diffusion equation.

We now study numerically Eq. \eqref{NPME} on randomly generated connected networks (Fig. \ref{fig:1}b-d). We start by placing a determined density on random nodes and we observe how this density spreads across the network. Interestingly the pattern that appears resembles that shown in Fig \ref{fig:1}a, as already happened for infinite graphs. While in the linear diffusion case (Fig. \ref{fig:1}b) all nodes have non-zero densities after arbitrarily small times, for the cases $m = 2,3$ (Fig. \ref{fig:1}c-d) some nodes seem to have zero density after positive times, thus showing a finite propagation speed. This is a major difference with respect to commonly used linear diffusion operators in network science.

This behaviour is easily understood when one looks at the relaxation time \eqref{eq:relaxation_time}. In the porous-medium case, the diffusion coefficient at the stationary distribution is given by Eq. \eqref{Diffusion_PME} and hence, the relaxation time is minimum for $m^* = -1/\log\Bar{\rho}$. In the figure we show values of the exponent satisfying $m>m^*\approx 0.22$, where $h(\Bar{\rho})$ is a decreasing function of $m$, thus explaining why larger values of the exponent yield slower diffusion.

\subsection{Comparing linear and nonlinear dynamics}

The differences between the linear and nonlinear cases can be further analyzed via a time rescaling of Eq. \eqref{NPME}. If we rescale this equation by the relaxation time given by Eq. \eqref{eq:relaxation_time}
\begin{equation}
        \frac{\mathrm{d}\rho_i}{\mathrm{d}t} = -{\lambda}{h(\Bar{\rho})^{-1}} \sum_jL_{ij}\left(f(\rho_j)g(\rho_i)-f(\rho_i)g(\rho_j)\right),\label{General_Model_rescaled}
\end{equation}
then for any choice of $f$ and $g$ we will have $\tau^{-1}_1 = \lambda\mu_1$. Now that the timescale of the process is decoupled from the chosen nonlinearities, one can easily compare several models. We focus again on the porous-medium case.

\begin{figure}
    \centering
    \includegraphics[width = .78\textwidth]{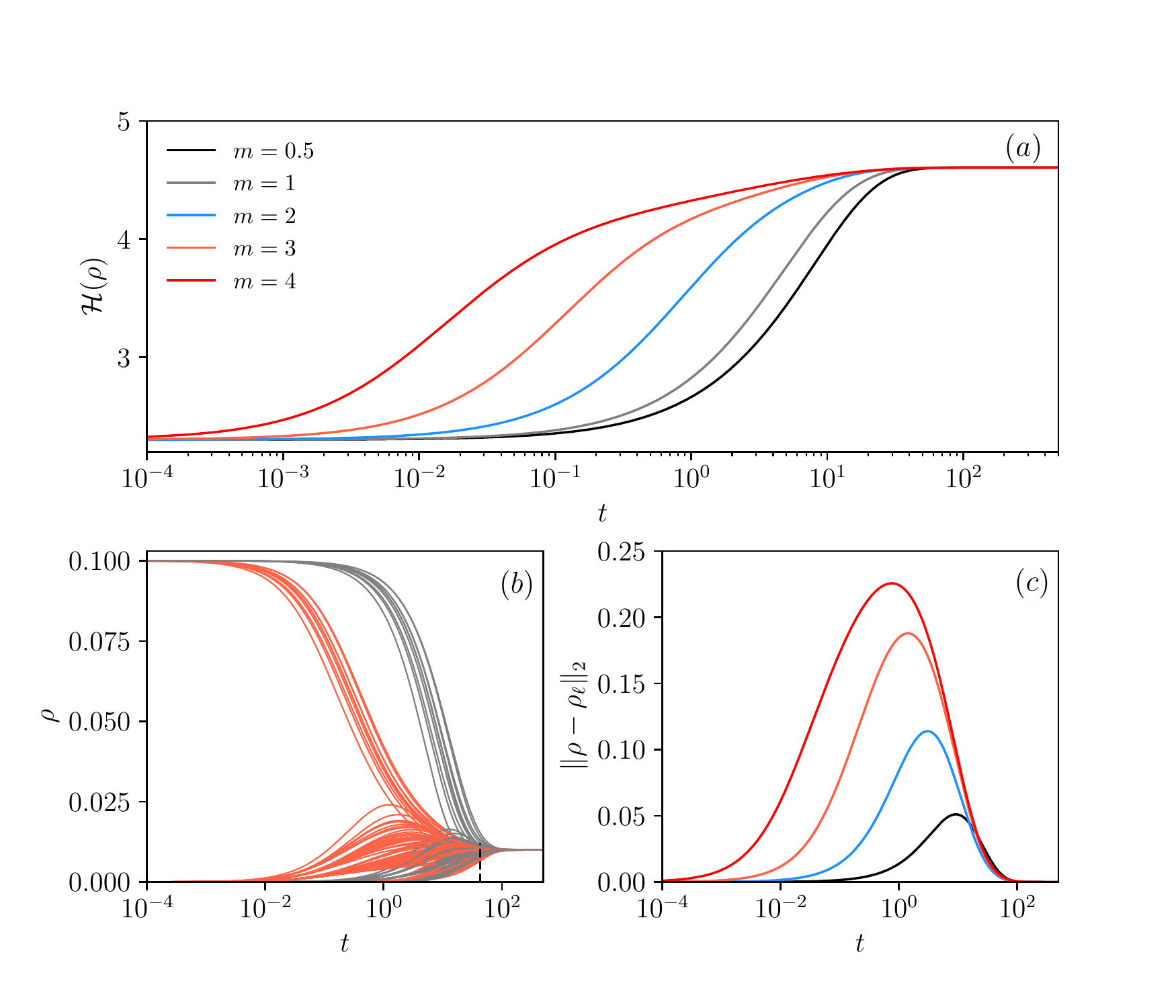}
    \caption{Comparing linear and nonlinear dynamics in the porous-medium case. The general model Eq. \eqref{General_Model} is rescaled by $h(\Bar{\rho})^{-1}$ in order to study the differences due solely to nonlinear effects. (a) Shannon entropy of $\rho$ given by Eq. \eqref{General_Model_rescaled} for different values of the exponent $m$. (b) Particular solutions for the linear case in gray ($m = 1$) and nonlinear case in red ($m = 3$). (c) Density difference for different exponents and the linear case given by $\rho_\ell$. Nonlinear diffusion shows very different dynamics from the linear case. All numerical simulations were performed in Erdos-Renyi networks with $M = 100$, $p = 0.1$ and using $\lambda = \Bar{\rho} =  0.01$. At time $t = 0$, we place a positive density at 10 randomly chosen nodes. All simulations share the same initial condition. }
    \label{fig:diversity}
\end{figure}

In order to evaluate the differences induced by different exponents $m$, we measure the Shannon entropy of $\rho$ as a function of time  $\mathcal{H}(\rho)=-\sum_i\rho_i\log\rho_i$ (Figure \ref{fig:diversity}a). Note that the entropy is maximum when $\rho$ is a uniform distribution and then this can be understood as a measure of the diversity of $\rho$. Other possible metrics, like the diversity index \cite{lambiotte_schaub_2022}, or different measures of variance on graphs \cite{devriendt2022variance}, give qualitatively similar results. In the figure, we observe how varying $m$ does indeed give place to different behaviours, with a higher level of uniformity obtained for larger values of the exponent.

These results are easily understood when one takes into account the speed of propagation of the process which, in the PDE setting, is infinite for the linear case and finite and decreasing with $m$ for $m>1$. On graphs, the situation is similar, and one obtains a faster decay of solutions for the linear case (and $m<1$). When compared with the nonlinear cases with slower diffusion, fixing the timescale means that a higher level of uniformity needs to be reached earlier in time -- Figure \ref{fig:diversity}b. This is accentuated the slower the diffusion mechanism or equivalently, the larger the exponent $m$ -- see the differences between the nonlinear cases and the linear case that we denote by $\rho_{\ell}$ in Figure \ref{fig:diversity}c.

\subsection{Timescale separation in modular networks}
From Eq. \eqref{eq:relaxation_time} we see that our previous arguments can actually be generalised to give the next slowest relaxation time of the process, $\tau_2$. Now $\tau_2^{-1}=h(\Bar{\rho})\lambda\mu_2$ is given by the second smallest non-zero eigenvalue of the graph Laplacian  $\mu_2$. This means that for times $t\gg \tau_2$, the network density is described by a linear combination of $u_1e^{-t/\tau_1}$, which is the leading term in the long-time limit, and also $u_2 e^{-t/\tau_2}$, where $u_2$ is the eigenvalue of $L$ associated with $\mu_2$.

\begin{figure}
    \centering
    \includegraphics[width = .48\textwidth]{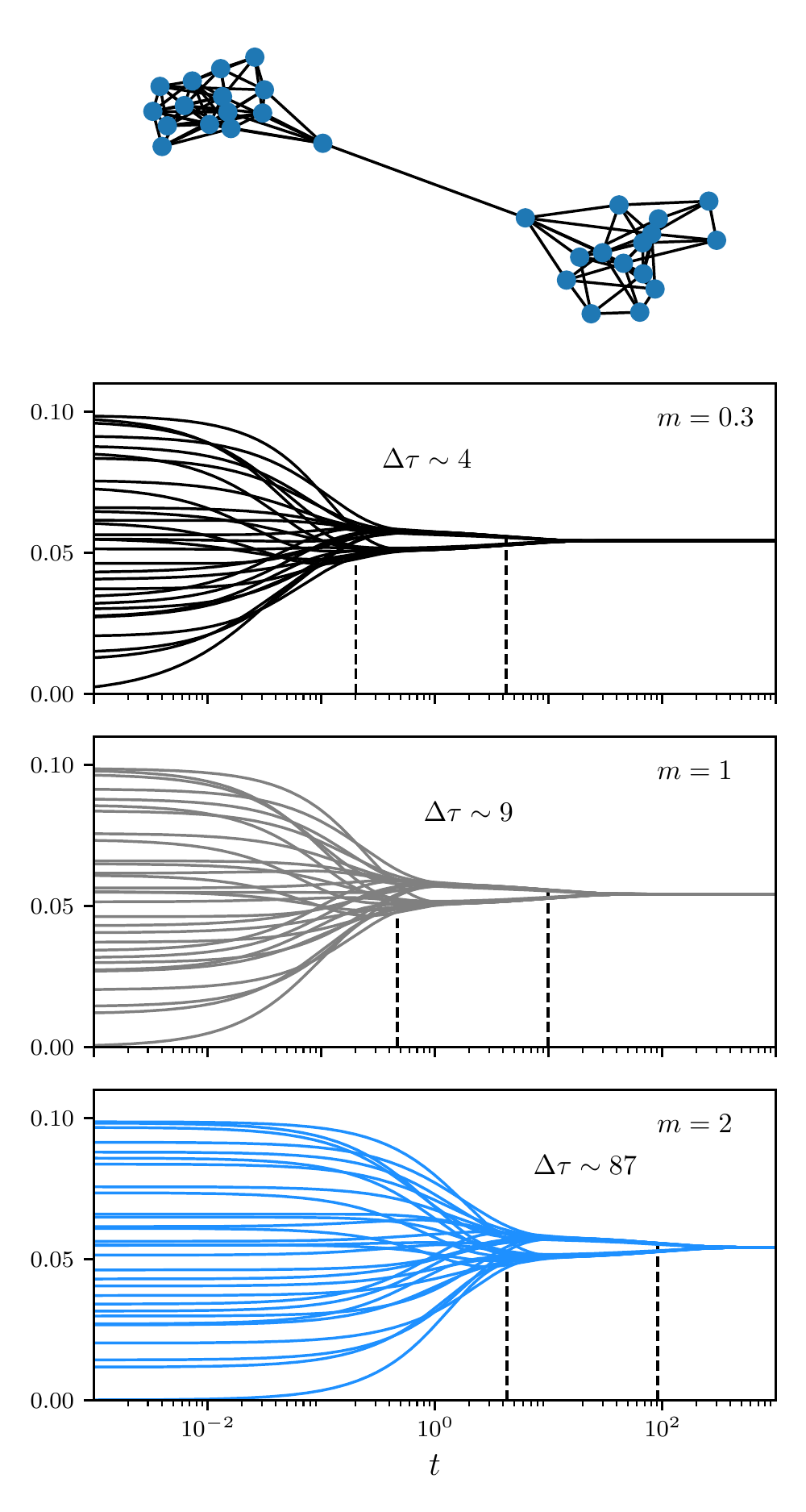}
    \caption{Nonlinear diffusion on modular networks. Here we use a network with two large components with $M = 15$ nodes each, and an edge that serves as a link. For every numerical simulation $\Bar{\rho} = 0.05$, $\lambda = 1$. Dashed lines represent the times $\tau_1$ and $\tau_2$.}
    \label{fig:sbm}
\end{figure}

In the case of graphs with a strong modular structure, the difference between $\mu_1$ and $\mu_2$ can be very large, thus giving place to two separated timescales, $\tau_1$ and $\tau_2$. During the first stages of the process, diffusion brings nodes in the same community to a state of local consensus, where they have similar densities, and after this first timescale, the different communities diffuse towards the global stationary state -- see Figure \ref{fig:sbm}. These considerations can be helpful in some applications because they allow for simplified representations of the network \cite{Schaub2016GraphPA}.

However, note that in general, the timescale separation depends also on the diffusion coefficient. As $\Delta\tau = \tau_1 - \tau_2\sim h(\Bar{\rho})^{-1}$, this means that $\Delta\tau$ could either be increased or decreased according to the nonlinearity. In particular, for the porous-medium case and small average density $\Bar{\rho}$, we obtain that $\Delta\tau$ increases with $m$ -- see Figure \ref{fig:sbm}.

Although these considerations hold even cases where the difference between $\mu_1$ and $\mu_2$ is small and hence no timescale separation is observed, one must be cautious when using random-walk dynamics to infer network modularity. In particular, for fast-diffusion equations -- for which $h(\rho)$ is large for small densities -- noisy measurements coupled with fast global dynamics might result  in community identifiability issues \cite{asllani2020dynamics}.
\section{Insights into reaction-diffusion systems}

As an application of the above ideas we briefly study reaction-diffusion equations of the type
\begin{equation}
    \frac{\mathrm{d}\rho_i}{\mathrm{d}t} = \left(\mathcal{L}\rho\right)_i + \alpha\rho_i(1-\rho_i),
    \label{eq:RD}
\end{equation}
where $\mathcal{L}$ is a diffusion operator, and the second term represents logistic growth with rate $\alpha$. We now consider a network where $\rho_i(0) = 0$ for $i\neq s$. At $t = 0$ we place a positive density $\rho_s(0)$ on a random node $s$, and study how the system evolves.

For general reaction-diffusion systems, the operator $\mathcal{L}$ could take many forms \cite{PropagationComplexNetworks, estradaNonlocal,diaz2022time}, but here we focus on the nonlinear operator given by Eq. \eqref{General_Model}. The resulting equation has recently been studied in the context of neurodegenerative diseases in \cite{FisherKPPGoriely,PutraGorielyNeuro} in the case where $\mathcal{L}$ represents linear diffusion -- $f(x) = x$, $g(x) = 1$. For extensions to two species with the same operator see for example \cite{AsllaniTuringPatterns,VanGorderRD}.

Before moving further, we remark that Eq. \eqref{eq:RD} constitutes a Fisher-KPP equation on networks. Probably the most interesting feature about this equation in the continuum setting is the formation of invasion fronts which propagate with a constant speed, and, studying the shape of the front as well as the speed of invasion, is  of particular interest  in mathematical biology. One could ask then if similar behaviour can be observed on networks and whether one can characterize these fronts.

This problem has been studied in simpler settings and only for linear diffusion, for example in the case of a lattice \cite{callaghan2006stochastic,bramson_lattice} as well as for homogeneous trees \cite{HoffmanFisherKPP,besse2021spreading}. In the lattice case, one always observes front propagation with a wave speed which depends on the scales of proliferation and diffusion. In the limit of large diffusion, one can derive that the speed tends to that of the linear diffusion Fisher-KPP PDE. On the other hand, in the case of homogeneous trees, this behaviour changes, and the speed of the front satisfies a non-monotonic relationship with the diffusion coefficient. In fact, above certain values of the diffusion coefficient, spreading might not occur and the solution converges to zero. In the limit of vanishing diffusion, the wave speed is decoupled from the diffusion coefficient, which predicts a constant speed value for more general networks.

Extending these results to the nonlinear diffusion setting seems challenging but it is certainly an interesting problem to explore. Here however, we focus on a qualitative study of Eq. \eqref{eq:RD}, and explore how considering nonlinear diffusion might influence mass propagation on random networks.

As already noted in \cite{FisherKPPGoriely}, Eq. \eqref{eq:RD} has two associated timescales $\tau_1$ and $\alpha^{-1}$, given by diffusion and growth, respectively. Depending on the relative strength of the two mechanisms, we can distinguish two regimes: a \emph{diffusion-dominated} regime, in which diffusion operates at much shorter timescales and thus $\tau_1 \ll \alpha^{-1}$; and a \emph{growth-dominated} regime where $\tau_1 \gg \alpha^{-1}$.
Note however that the mean-field approximation used to derive Eq. \eqref{General_Model} might be worse in this last regime where proliferation dominates.

\begin{figure}
    \centering
    \includegraphics[width = .7\textwidth]{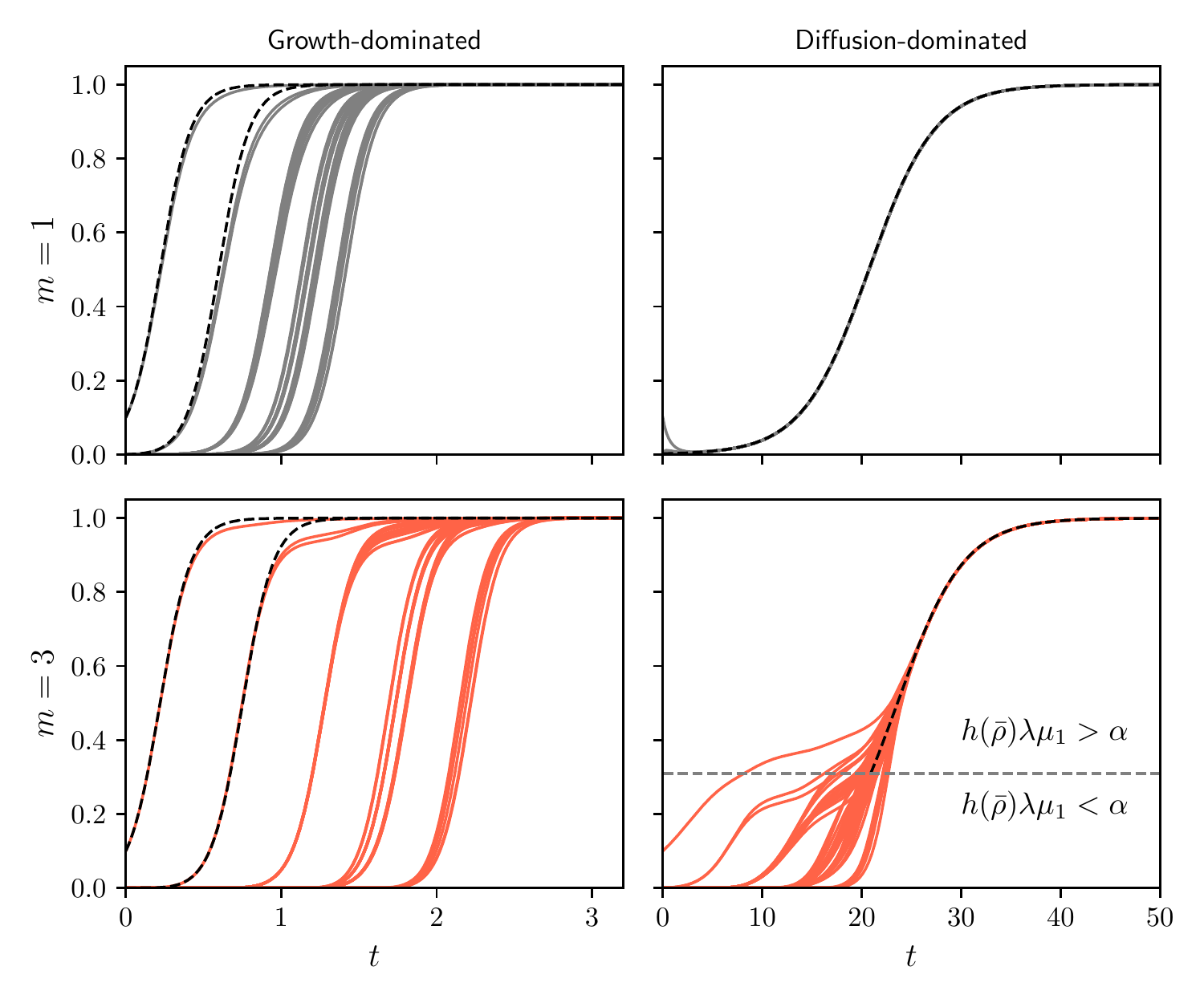}
    \caption{ A Fisher-KPP type equation in the growth-dominated regime (left, $\lambda = 0.05$, $\alpha = 10$) and the diffusion-dominated regime (right, $\lambda = 1$, $\alpha = .3$): comparing  linear (top) with nonlinear diffusion with exponent  $m =3$ (bottom). Simulations take place on Erdos-Renyi networks with a number of nodes $M = 50$ and a probability of drawing any possible edge $p = 0.1$. Black dashed lines correspond to the asymptotic solutions derived in Appendix \ref{sec:asymptotic}. Gray dashed line is given by \eqref{eq:crit_density}.}
  \label{fig:Reaction_Diffusion}\end{figure}

For linear diffusion, these two regimes are easily characterized, as diffusion is density-independent, and depends directly on the quotient $\lambda\mu_1/\alpha$. The two regimes can be seen in Figure \ref{fig:Reaction_Diffusion} in gray lines. In the growth-dominated regime, proliferation occurs at a much larger rate than diffusion, causing nodes closer to the initial seed to reach higher densities faster. However, the densities at nodes that are further from the initial seed, are close to zero for times smaller than $\lambda\mu_1\ll \alpha$. On the other hand, in the diffusion-dominated case where $\lambda\mu_1/\alpha\gg 1$, all nodes rapidly reach the consensus state via diffusion, and then grow together following logistic growth.

This behaviour is somewhat different in the nonlinear case, as the timescale of diffusion is density-dependent $\tau_1\sim h(\Bar{\rho})^{-1}$, with $\Bar{\rho}$ varying in time due to the presence of proliferation. Note now, that the two regimes are given instead by the quotient $h(\Bar{\rho})\lambda\mu_1/\alpha$, and hence it is possible to move from one regime to the other one in time, depending on the evolution of the average density $\Bar{\rho}$. For the porous-medium equation, we see that diffusion is always weak at low densities, where $\Bar{\rho}$ is small. In the case of a large proliferation rate, the timescale separation between nodes becomes clearer, and nodes with the same step distance to the initial seed tend to cluster and grow together to the steady state. In the opposite case, however, diffusion can still be weaker for early times even with a small proliferation rate. As proliferation occurs, diffusion becomes more and more important until it overcomes growth. This is shown in Figure \ref{fig:Reaction_Diffusion} for $m = 3$. The critical density $\Bar{\rho}_c$ at which this regime transition happens can be found, and satisfies $h(\Bar{\rho}_{c})\lambda\mu_1 = \alpha$. In the porous-medium case
\begin{equation}
    \Bar{\rho}_{c} = \left(\frac{\alpha}{m\lambda\mu_1}\right)^{1/(m-1)}\label{eq:crit_density}.
\end{equation}

For some applications, including the study of neurodegenerative diseases \cite{FisherKPPGoriely}, it is instructive to analyze the time needed to reach certain densities at every node. For that purpose, one can construct asymptotic solutions in each one of the discussed regimes which approximate very well the evolution of the density at each node -- see black dashed lines in Figure \ref{fig:Reaction_Diffusion}. Although inspired by \cite{FisherKPPGoriely}, we explain in Appendix \ref{sec:asymptotic} the details of how to find these approximations.

\section{Conclusions and outlook}

To summarize, here we have provided a new link between diffusion PDEs and random-walk models on networks. This gives a new interpretation for previous works developed at the scale of individual agents \cite{CarlettiNonlinear,AsllaniCrowding}. Moreover being able to build the connection with the macroscopic PME and similar equations helps to understand why nonlinear diffusion on networks yields mass propagation at a finite speed, in contrast with the commonly used linear diffusion operator. Generalisations of this work could include accounting for different interacting species \cite{twoSpeciesNetworks} or for further interaction terms in the form of an external or interaction potential \cite{EspositoNonlocal,MarkusTwoSpecies}.

Some of the examples considered here also provide with a useful framework for applications. For instance, and as suggested before \cite{CarlettiNonlinear}, \emph{fast diffusion} can be thought as an efficient way of network exploration, and on the other hand, \emph{slow diffusion} may emerge in applications where individuals motility is reduced due to crowding effects. In other contexts, it would also be interesting to explore the validity of ignoring crowding effects and describing movement as simple linear diffusion.

\begin{acknowledgments}
The author wishes to thank Ruth Baker, José A. Carrillo and Renaud Lambiotte for helpful discussions and  suggesting significant improvements to earlier versions of this manuscript.
This work was made possible through the support of a fellowship from "la Caixa" Foundation (ID 100010434) with code LCF/BQ/EU21/11890128.
\end{acknowledgments}

\appendix

\section{Mean-field limit from system-size expansion}
Here we derive the mean-field equation Eq. \eqref{General_Model} from the master equation Eq \eqref{eq:master_equation}, using a system-size expansion. The size of the system in our case, is given by the carrying capacity $K$, and thus, $K\rightarrow\infty$ corresponds to the thermodynamic limit. We follow \cite[Chapter X]{VanKampen} closely.

The basic idea behind the method is to take advantage of the structure of the transition probabilities \eqref{eq:Transition_edge_centric} in order to find an expansion of the master equation in terms of powers of $K$. Note that when $K$ is large, the transition probabilities \eqref{eq:Transition_edge_centric} are small, and we can expect  $P(n_i,t)$ to be peaked around the macroscopic mean $n_i\sim K \rho_i$, with a width of order $K^{1/2}$. Under this setting, $n_i$ follows
\begin{equation}
    n_i = K\rho_i + K^{1/2}\xi_i,
\end{equation}
with $\xi_i$ giving the fluctuations in the number of particles at node $i$. Now the transition probabilities may be written for large $K$ as
\begin{align}
    T(n_i-1,n_j +1|n_i,n_j) &= \lambda A_{ij}f\left(\frac{K \rho_i + K^{1/2}\xi_i}{K}\right)g\left(\frac{K \rho_j + K^{1/2}\xi_j}{K}\right)\nonumber \\& = \lambda A_{ij}f(\rho_i)g(\rho_j) + O(K^{-1/2}).\label{eq:transition_expansion}
\end{align}

We are interested in studying $\rho_i=\langle n_i\rangle/K$ for large $K$, where $\langle n_i\rangle = \sum_{\mathbf{n}}n_iP(\mathbf{n},t)$. By taking the time derivative of $\langle n_i\rangle$ and using the differential equation for the master equation \eqref{eq:master_equation}, we obtain -- see also \cite{CarlettiNonlinear} --
\begin{equation}
    \frac{\mathrm{d}}{\mathrm{d}t}\langle n_i\rangle = -\sum_j\left[\langle T(n_i-1,n_j+1|n_i,n_j)\rangle-\langle T(n_i+1,n_j-1|n_i,n_j)\rangle\right].
\end{equation}
Now rescale time so that $t\mapsto t/K$ and use Eq. \eqref{eq:transition_expansion} in the above expression to find, to leading order in $K$, the mean-field equation
\begin{equation}
     \frac{\mathrm{d}\rho_i}{\mathrm{d}t} = -\lambda\sum_j \left[A_{ij}f(\rho_i)g(\rho_j)-A_{ji}f(\rho_j)g(\rho_i)\right],
\end{equation}
which might be rearranged into Eq. \eqref{General_Model} by using $A_{ij}=A_{ji}$.

\section{A porous-medium type equation in the node-centric setting}
\label{appendix:edge_centric}

Here we study Eq. \eqref{General_Model} in the node-centric setting, as it was initially derived in \cite{CarlettiNonlinear}. In this case the transition probabilities read
 \begin{equation}
 T(n_i-1,n_j +1|n_i,n_j) = \frac{\lambda A_{ij}}{k_i}f\left(\frac{n_i}{K}\right)g\left(\frac{n_j}{K}\right).
 \label{eq:Transition_node_centric}
 \end{equation}
 These give place, via a mean-field approximation, to a similar equation for the evolution of the node density $\rho_i = \lim_{K\rightarrow\infty}\langle n_i\rangle/K$
 \begin{equation}
       \frac{\mathrm{d}\rho_i}{\mathrm{d}t} = -\lambda \sum_jL'_{ij}\left(\frac{k_i}{k_j}f(\rho_j)g(\rho_i)-f(\rho_i)g(\rho_j)\right),\label{General_Model_node_centric}
\end{equation}
where now instead we have the random-walk normalized Laplacian defined by $L_{ij}'  = \delta_{ij} - A_{ij}/k_i$. Note that in the case of regular networks where $k_i = k$, the above expression and Eq. \eqref{General_Model} are essentially the same equation, up to a constant given by the network degree. This means that Eq. \eqref{General_Model_node_centric} is also valid for deriving the same macroscopic limits that we discussed in Section \ref{macro_limit}.

It was already proven in \cite{CarlettiNonlinear} that the stationary distribution of \eqref{General_Model} satisfies the relation
\begin{equation}
    \frac{f(\rho_i^*)}{k_ig(\rho_i^*)}=  \frac{f(\rho_j^*)}{k_jg(\rho_j^*)}=c,
\end{equation}
for a given constant $c$ that can be found using conservation of mass. In the local repulsion model -- i.e. $f(x) = x^m$ and $g(x) = 1$ -- the stationary distribution reads
\begin{equation}
  \rho_i^* =  ck_i^{1/m},
\end{equation}
where $c = \sum_i\rho_i/\sum_i k_i^{1/m}= \Bar{\rho}/\langle k^{1/m}\rangle$. It is informative to study how the stationary distribution depends on $m$. In the limit of large exponents $m$, $k_i^{1/m}\rightarrow 1$ and thus the dependence on the node degree is erased. In this case we can approximate
\begin{equation}
    \rho_i^*\approx \Bar{\rho},\quad\text{as } m\rightarrow\infty,
\end{equation}
obtaining that in the stationary state, all nodes have the same density of walkers. Note that in this limit, we obtain the same stationary distribution as in the edge-centric case. As we will see, the relaxation time also tends to the one obtained in the main text.

On the other hand, a completely different phenomenon occurs in the limit where $m$ tends to zero. In this case, walkers tend to accumulate in nodes with higher connectivity, and in the limit $m\rightarrow 0$, only nodes with maximum degree have non-zero densities. To see that, take a particular node $i$. If $k_i<k_{\text{max}}$, where  $k_{\text{max}}$ is the maximum degree in the network, then $k_i^{1/m}/\sum_i k_i^{1/m}\rightarrow 0 $. However, if $k_i = k_{\text{max}}$, we have $k_i^{1/m}/\sum_i k_i^{1/m}\rightarrow 1$. Then we have found
\begin{equation}
    \lim_{m\rightarrow 0^+}\rho_i^* = \left\{
\begin{array}{lcl}
\frac{\Bar{\rho} M}{\# k_{\text{max}}}
&\mbox{for} & k_i = k_{\text{max}}, \\
0
&\mbox{for} & k_i \neq k_{\text{max}}, \\
\end{array}
\right.
\end{equation}
where  $\# k_{\text{max}}$ is the number of nodes with maximum degree. We see then that the spread of the stationary distribution goes from $\Bar{\rho} M/\# k_{\text{max}}$ in the $m\rightarrow 0$ limit, to zero when $m\rightarrow \infty$. 

Obtaining a closed-form expression for the relaxation time of Eq. \eqref{General_Model_node_centric} seems more challenging than in the edge-centric case and maybe not very illuminating. However, progress can be made in the porous-medium case again. Again, we perturb around the steady state and linearize the resulting equation.

We thus seek for solutions of the form $\rho_i = \rho_i^* + \epsilon  \eta_ik_i^{1/m}$ for $\epsilon \ll 1$. The scaling in the perturbation is only chosen in order to facilitate the notation in the following arguments. Bringing together the $O(1)$ terms gives the stationary solution. Up to order $O(\epsilon)$ we have
\begin{equation}
\frac{\mathrm{d}\eta}{\mathrm{d}t}=-mc^{m-1}D_m L\eta :=-mc^{m-1}L_m\eta,
\label{rho_tilde_node_centric}
\end{equation}
where $D_m = \text{diag}\left(k_1^{-1/m},\ldots,k_N^{-1/m}\right)$ and $L_m$ is defined by the rightmost identity. Note that in the linear diffusion case ($m=1$) $D_m = D^{-1}$ and $L_1 = L'$ is the random-walk normalized Laplacian.

In order to find the relaxation time we follow a standard procedure \cite{LambiottePorterMasuda}, and project $\eta$ onto a basis of eigenvectors of $L_m$. In the long-time limit, the solution can be written approximately in terms of the eigenvector with the smallest non-zero eigenvalue,  $\tilde{\mu}_1$. Thus we find that
\begin{equation}
    \tau_1^{-1} = mc^{m-1}\tilde{\mu}_{1}.
    \label{relaxation_time_node_centric}
\end{equation}
Note here that $\tilde{\mu}_{1}$ also depends on $m$.
Again, the relaxation time seems to be linked to the diffusion coefficient of the PME Eq. \eqref{Diffusion_PME} as we can write $\tau_1^{-1}\sim h(c)$. This is the same scaling that we found in the edge-centric case with the substitution $\Bar{\rho}\mapsto \Bar{\rho}/\langle k^{1/m}\rangle$ in Eq. \eqref{eq:relaxation_time}.

Again, we explore how this relaxation time is related to the edge-centric case, given by \eqref{eq:relaxation_time}. First of all, note that using a very naive mean-field approximation which accounts for neglecting network structure $k\mapsto\langle k\rangle$, would give a relaxation time equivalent to the one found in the main text
\begin{equation}
     \tau_{MF}^{-1} = \frac{m\Bar{\rho}^{m-1}\mu_1}{\langle k\rangle}.
     \label{MF_relaxation_time}
\end{equation}
This expression is in fact exact for regular networks.

In the slow diffusion limit, one can find a very similar relationship suggesting again the scaling $\tau_1^{-1}\sim h(\Bar{\rho})$. In the large $m$ limit, note that the matrix $D_m$ tends to an identity matrix. In fact, one may write for every entry $k_i^{1/m} = 1 + \frac{\log k_i}{m} + O(m^{-2})$, and thus
\begin{equation}
     L_m = D_m L = L-\frac{1}{m}\tilde{D}L +O(m^{-2}),
\end{equation}
with $\tilde{D} = \text{diag}\left(\log k_1,\ldots,\log k_M\right)$.
Now we can use a standard perturbation theory argument and expand $\tilde{\mu}_1 = \mu_1 + O(m^{-1})$. The first order correction can actually be found in terms of the above expansion, but the leading order is sufficient for our purposes.

In the large $m$ limit, $c^{m-1}$ can be expanded similarly
    \begin{eqnarray}c^{m-1}= \Bar{\rho}^{m-1}e^{-\langle\log k\rangle}\left(1+O(m^{-1})\right).
          \label{c_approx}
     \end{eqnarray}
     Then, together with Eq. \eqref{relaxation_time_node_centric}, we have obtained
     \begin{equation}
  \tau_1^{-1} = \frac{m\Bar{\rho}^{m-1}\mu_1}{\left(k_1\ldots k_M\right)^{1/M}}\left(1+O(m^{-1})\right),\label{relaxation_time_approx} \end{equation}
  which is essentially the same equation as \eqref{MF_relaxation_time}, substituting the mean degree $\langle k\rangle$ by its geometric mean.

Although obtained through different approximations, Eq. \eqref{relaxation_time_approx} and Eq. \eqref{MF_relaxation_time} show the same dependence of $\tau_1$ with respect to the exponent $m$. This suggests that $\tau_1$ also follows the scaling derived in the main text for  the edge-centric and macroscopic cases: $\tau_1^{-1} \sim h(\Bar{\rho})$.

\section{Asymptotic solutions for Fisher-KPP type equations}
\label{sec:asymptotic}

We discuss here how to obtain approximations for the solutions of Eq. \eqref{eq:RD}, in the case given by the PME $f(x) = x^m$, $g(x) = 1$. We distinguish between the two mentioned regimes.

The diffusion-dominated regime is easier, since it can be described by the solution of the usual logistic equation. In such case, the densities at each node grow simultaneously after some time $t_c$ determined by the condition $\langle\rho\rangle = \Bar{\rho}_c$. For $t>t_c$ thus, all nodes have similar densities and we can ignore diffusion and write \begin{equation}
  \frac{\mathrm{d}\rho_i}{\mathrm{d}t} = \alpha\rho_i(1-\rho_i),\quad\rho_i(t_c) = \Bar{\rho}_c,
\end{equation}
with solution
\begin{equation}
\rho_i(t) = \frac{\Bar{\rho}_ce^{\alpha (t-t_c)}}{1 + \Bar{\rho}_c\left(e^{\alpha (t-t_c)} - 1\right)}.
    \label{eq:asymptotic_diff_dominated}
\end{equation}
This curves are plotted in Figure \ref{fig:Reaction_Diffusion} in black dashed lines for the diffusion-dominated case. Note that for linear diffusion we can approximate $t_c\approx 0$.

In the growth-dominated case one can expect $\lambda$ to be small in comparision with $\alpha$, and then it makes sense to consider an asymptotic solution in terms of powers of $\lambda$. Here we follow an asymptotic method developed in \cite{VariationParameters, FisherKPPGoriely} -- see the latter for other possibilities. Whenever $\lambda = 0$, the solution takes essentially, the same form as in Eq. \eqref{eq:asymptotic_diff_dominated}. Without imposing initial conditions we have
\begin{equation}
\rho_i(t) = \frac{K_i e^{\alpha t}}{1 + K_i\left(e^{\alpha t} - 1\right)}.
    \label{eq:asymptotic_growth_dominated}
\end{equation}
Now, motivated by the numerical simulations shown in Figure \ref{fig:Reaction_Diffusion}, we consider the ansatz: $K_i = K_i^{(0)} + \lambda^{\gamma_i}\tilde{K}_i$, where $\gamma_i$ is the distance between node $i$ and the initial seed. We can now solve iteratively for different orders in $\lambda$. The solution up to order $\lambda^0$, is $K_s^{(0)} = \rho_s(0)$ and $K_i^{(0)} = 0$ for $i\neq s$.

Looking at the $O(\lambda)$ terms, we obtain the solution for nodes that are at distance $\gamma_i = 1$ from the initial seed. This yields the differential equation for $\tilde{K}_i$
\begin{equation}
    \frac{\mathrm{d}\tilde{K}_i}{\mathrm{d}t} = -L_{is}e^{-\alpha t}\rho_s(t)^m,\quad \tilde{K_i}(0) = 0;
\end{equation}
which can be integrated to give
\begin{eqnarray}
    \tilde{K}_i(t) = -\frac{L_{is}\rho_s(0)^m}{\alpha(m-1)(1-\rho_s(0))}
    \left[1-\left(\frac{\rho_s(t)}{\rho_s(0)}\right)^m\left(\rho_s(0) + \left(1 - \rho_s(0)\right)e^{-\alpha t}\right)\right].&
\end{eqnarray}

In the linear diffusion case ($m = 1$), we have instead
\begin{equation}
    \tilde{K}_i(t) = -\frac{L_{is}\rho_s(0)}{\alpha(1-\rho_s(0))}\left[\alpha t-\log\left(\rho_s(0)(e^{\alpha t}-1) + 1\right)\right].
    \end{equation}

    These asymptotic solutions are represented in Figure \ref{fig:Reaction_Diffusion}. Note that there is a very good agreement between the numerical solution and the derived approximations. When densities are close to $1$, these might fail in the nonlinear case due to the increase of the diffusion coefficient with the density.

    The solutions at distance $\gamma_i>1$ from $s$ can be found iteratively in terms of these solutions. For more details, we refer the reader to \cite{FisherKPPGoriely}.

\end{document}